\documentstyle[hep93]{article}
\def\lsim{\mathrel{\rlap {\raise.5ex\hbox{$ < $}}
{\lower.5ex\hbox{$\sim$}}}}
\def\gsim{\mathrel{\rlap {\raise.5ex\hbox{$ > $}}
{\lower.5ex\hbox{$\sim$}}}}

\newcommand{\pr}{\paragraph{}}
\newcommand{\be}{\begin{equation}}
\newcommand{\ee}{\end{equation}}
\newcommand{\bea}{\begin{eqnarray}}
\newcommand{\nn}{\nonumber}
\newcommand{\eea}{\end{eqnarray}}

\def\gappeq{\mathrel{\rlap {\raise.5ex\hbox{$>$}}
{\lower.5ex\hbox{$\sim$}}}}
 
\def\lappeq{\mathrel{\rlap{\raise.5ex\hbox{$<$}}
{\lower.5ex\hbox{$\sim$}}}}
 
\begin{document}
\begin{flushright}
OUTP-96-28P \\
gr-qc/9606008\\
\end{flushright}
 
\title{Eluding the No-Hair Conjecture for Black Holes
\thanks{Invited
Contribution to the
5th Hellenic School and Workshops on High-Energy Physics,
Kerkyra/Corfu (Greece), 3-24 September 1995}
\\}

\author{N. E. Mavromatos \\
        {\em University of Oxford, Dept. of Theoretical Physics
(P.P.A.R.C. Advanced Fellow), 1 Keble Road, OX1 3NP Oxford, U.K. }
         }
\abstract{
I discuss a recent analytic proof of bypassing
the no-hair conjecture for two interesting
(and quite generic) cases of four-dimensional black holes:
(i) black holes
in Einstein-Yang-Mills-Higgs (EYMH) systems and
(ii) black holes in higher-curvature (Gauss-Bonnet (GB) type)
string-inspired gravity.
Both systems are known to possess black-hole
solutions with non-trivial scalar hair outside the horizon.
The `spirit' of the no-hair conjecture, however, seems to be
maintained either because the black holes are unstable (EYMH),
or because
the hair is of secondary type (GB), i.e. it does not lead to
new conserved quantum numbers.
}
\maketitle
\section{Introduction}
\pr
The `no-hair conjecture' of Black holes may be best atttributed
to Wheeler~\cite{wheeler}, who, motivated by earlier uniqueness
theorems on Black-hole solutions of Einstein-Maxwell
theory~\cite{israel},
stated that {\it all exteriors of stationary black hole solutions
are uniquely characterized by at most three conserved `charges':
mass, angular momentum and electric charge}, i.e there are no
non-trivial exterior fields, outside the horizon of a black-hole,
other than those associated with long range
Abelian gauge forces, obeying a
Gauss law constraint.
\pr
The surprising discovery of the Bartnik-McKinnon (BM) non-trivial particle-like
structure~\cite{bartnik} in the Einstein-Yang-Mills system
opened many possibilities for the existence of non-trivial
solutions to Einstein-non-Abelian-gauge systems~\cite{smoller}.
Indeed, soon after its discovery, many other
self-gravitating structures with non-Abelian gauge fields
have been discovered~\cite{selfgrav}. These
include black holes
with non-trivial non-Abelian
hair, thereby leading to the possibility of
evading the no-hair conjecture~\cite{wheeler,nohair}.
The physical reason for the existence of these classical
solutions is the `balance' between
the non-Abelian gauge-field repulsion
and the gravitational attraction.
As I shall argue in this talk~\cite{mw}, such a balance
allows for dressing black hole solutions
by non-trivial configurations (outside the horizon)
not only of non-Abelian gauge fields, whose presence
should not have come as a great surprise since they obey
the Gauss-law constraint in a rather similar spirit with the
electromagnetic-field case, but also
of (scalar) fields that are not associated with a Gauss-law.
It is the presence of the latter
that leads to an `apparent' evasion of the no-hair 
conjecture~\footnote{In this article I shall concentrate 
on classical hair. Quantum hair is not covered by the no-hair 
conjecture, and it is a totally different, but equally 
important, issue of black-hole physics, which, however, 
will not be touched upon here.}.
\pr
In this talk I will explain in some detail the
reasons for such a bypassing of the no-hair conjecture~\cite{mw},
by concentrating on two physically interesting
examples, which capture the essential generic features
of the problem.
The first example is
that of a spontaneously broken Yang-Mills theory
in a non-trivial black-hole space-time (EYMH)~\cite{greene}.
This system has been recently examined from a stability point of view,
and found to possess instabilities~\cite{winst,mw},
thereby making the physical importance
of the solution rather marginal, but also indicating another
dimension of the no-hair conjecture, not considered in the original
analysis, that of stability.
The fact that the hairy black hole is unstable implies a
violation of the `letter' but not of the `spirit' of the
no-hair conjecture. The second example is that of higher-curvature
string-inspired gravity, and in particular a dilaton-graviton
(super)string-effective action containing curvature-square terms
of Gauss-Bonnet (GB) type~\cite{rizos}.
The dilaton hair in that case is of secondary type~\cite{coleman},
i.e. it does not lead to new conserved quantum
numbers as it is expressed in terms of the $ADM$ mass~\cite{rizos}.
Again, the `spirit' of the no-hair conjecture seems to be maintained.
\pr
I will start my talk by briefly presenting a method, developed
in collaboration with E. Winstanley~\cite{mw},
for proving the possibility of an evasion of the no-hair
conjecture for these systems. Our approach was inspired
by
a recent
elegant proof
of the no-hair theorem for black holes by Bekenstein~\cite{bekmod}
in
a variety of cases with scalar fields~\cite{bekmod}. The theorem
is formulated in such a way so as to rule out a multicomponent
scalar field dressing an
asymptotically flat, static, spherically-symmetric
black hole. The basic
assumption of the theorem
is that the scalar
field is minimally coupled to gravity and bears a non-negative energy density
as seen by any observer, and the proof relies on very general principles,
such as energy-momentum conservation and the Einstein equations.
{}From the positivity assumption and the conservation equations
for the energy momentum tensor $T_{MN}$ of the theory, $\nabla ^M T_{MN}= 0$,
one obtains for a spherically-symmetric space-time background
the condition that near the horizon the radial component
of the energy-momentum tensor and its first derivative are negative
\be
    T _r^r < 0,  \qquad (T_r^r)' < 0
\label{zero}
\ee
with the prime denoting differentiation with respect to $r$.
This
implies that in such systems there must be regions in space outside
the horizon where both quantities in (\ref{zero}) change sign.
This contradicts the results following from Einstein's equations
though~\cite{bekmod}, and this {\it contradiction} constitutes
the proof of the no-hair theorem, since the only allowed non-trivial
configurations are Schwarzschild black holes.
We note, in passing,
that there are known exceptions to the original version of the
no-hair theorem~\cite{nohair}, such as conformal
scalar fields coupled to gravity, which
come from the fact that in such theories the scalar fields
diverge at the horizon of the black hole~\cite{confdiv},
and therefore
the original assumptions of the theorem are violated.
\pr
The interest for our case
is that the theorem, if true for the EYMH system, would
rule  out the existence
of non-trivial hair due to a Higgs field with a double (or multiple) well
potential, as is the case for spontaneous symmetry breaking.
Given that stability issues are not involved in the proof,
it would be
of interest to reconcile the results of the theorem with the situation
in the case of EYMH or GB systems,
where at least
we know that explicit black-hole solutions with non-trivial
hair exist~\cite{greene,rizos}.
As we shall
show below, the formal reason
for bypassing
the modern version
of the no-hair theorem~\cite{bekmod}
lies in the violation
of the key relation among the
components of the stress tensor, $T_t^t =T_\theta^\theta$,
shown to hold in the case of ref. \cite{bekmod}.
The physical reason
for the `elusion' of the above no-hair conjecture lies in the fact that
the presence of the repulsive non-Abelian gauge interactions,
or the higher-curvature gravitational terms,
balances the gravitational attraction, by producing terms that
make the result (\ref{zero}) inapplicable in this case.
For the GB system
there is an additional simpler reason to expect that
the no-scalar-hair theorem  can be bypassed.
In the presence of curvature-squared terms,
the modified Einstein's equation leads to an effective
stress tensor that involves the gravitational field.
This implies that the assumption of
positive definiteness of the time-component of this tensor,
which in the Einstein case is the local energy density
of the field,
breaks down,
leading to a relaxation of one of the key constraints/assumptions
of the no-hair theorem of ref. \cite{bekmod}.
 
\section{Eluding the no-hair conjecture for
the EYMH system}
\pr
We start our discussion from the EYMH system.
Consider the EYMH theory with a (gauge-fixed) Lagrangian
\be
{\cal L}_{EYMH} = -\frac{1}{4\pi} \left\{ \frac{1}{4} |F_{MN}|^2
+ \frac{1}{8} \phi ^2 |A_M|^2 + \frac{1}{2} |\partial _M \phi |^2
+ V(\phi ) \right\}
\label{onea}
\ee
where $A_M$ denotes the Yang-Mills field, $F_{MN}$ its field strength,
$\phi $ is the Higgs field and $V (\phi )$ its potential.
All the indices are contracted with the help of the
background gravitational tensor $g_{MN}$.
In the spirit of Bekenstein's
modern version of the no-hair theorem~\cite{bekmod}, we
now examine
the energy-momentum tensor of the model (\ref{onea}). It can be written in the
form
\be
 8 \pi T_{MN} = -{\cal E } g_{MN} + \frac{1}{4\pi} \left\{ F_{MP}F_{N}{}^{P}
+ \frac{\phi ^2}{4} A_M A_N + \partial _M \phi \partial _N \phi \right\}
\label{threea}
\ee
with ${\cal E} \equiv -{\cal L}_{EYMH} $.
We consider Yang-Mills fields of the form~\cite{greene}
\be
 A = (1 + \omega (r) ) [-{\hat \tau} _\phi d\theta + {\hat \tau}_\theta
\sin \theta d\varphi ]
\label{twelvea}
\ee
where $\tau _i $, $i =r,\theta,\varphi $
are the generators of the $SU(2)$ group
in spherical-polar coordinates.
\pr
Consider, now, an observer moving with a four-velocity $u^M $ . The observer
sees a local energy density
\be
\rho  = {\cal E} + \frac{1}{4\pi} \left\{ u^M F_{MP} F_N{}^{P} u^N +
\frac{\phi ^2}{4} (u^M A_M )^2 + (u^M \partial _M \phi )^2 \right\},
\qquad u^M u_M = -1.
\label{foura}
\ee
 
To simplify the situation let us
consider a space-time with a time-like killing vector, and suppose that the
observer moves along this killing vector. Then
$ u^M \partial _M \phi = 0 $ and by an appropriate gauge choice
$u^M A_M =0 = u^M F_{MN} $. This gauge choice is compatible with the
spherically-symmetric ansatz
for $A_M$ of ref. \cite{greene}. Under these assumptions,
\be
    \rho = {\cal E}
\label{fivea}
\ee
and the requirement that the local energy density as seen by any observer
is non-negative
implies
\be
{\cal E } \ge 0.
\label{sixa}
\ee
\pr
We are now in position to proceed with the announced proof of the bypassing
of the no-hair theorem of ref. \cite{bekmod} for the EYMH black hole of
ref. \cite{greene}.
To this end we consider a spherically-symmetric ansatz for the
space-time metric $g_{MN}$, with an invariant line element of the form
\be
  ds^2 = - e^{\Gamma } dt^2 + e^\Lambda dr^2
+ r^2 (d\theta ^2 + \sin^2 \theta d\varphi ^2),
\qquad \Gamma = \Gamma (r),~\Lambda = \Lambda (r).
\label{sevena}
\ee
To make the connection with the black hole case we further assume that the
space-time is asymptotically flat.
\pr
{}From the conservation of the energy-momentum tensor, following from the
invariance of the effective action under general co-ordinate transformations,
one has  for the $r$-component of the conservation equation
\be
        [(-g)^{\frac{1}{2}} T_r^r ]' - \frac{1}{2} (-g)^{\frac{1}{2}}
\left( \frac{\partial }{\partial r} g_{MN}\right) T^{MN} = 0
\label{eighta}
\ee
with the prime denoting differentiation with respect to $r$. The spherical
symmetry of the space time implies that $T_\theta ^\theta =
T_\varphi ^\varphi $.
Hence, (\ref{eighta}) can be written as
\be
   (e^{\frac{\Gamma+\Lambda}{2}} r^2 T_r^r )' - \frac{1}{2}
e^{\frac{\Gamma + \Lambda}{2}} r^2 \left[ \Gamma' T_t^t + \Lambda ' T_r^r
+ \frac{4}{r}
T_\theta ^\theta \right] = 0.
\label{ninea}
\ee
Observing that the terms containing $\Lambda $ cancel,
and integrating over the radial coordinate $r$ from the horizon
$r_h$ to a generic distance $r$, one obtains
\be
   T_r^r (r) = \frac{e^{-\frac{\Gamma}{2}}}{2 r^2}
\int _{r_h}^r dr e^{\frac{\Gamma}{2}} r^2 \left[ \Gamma' T_t^t +
\frac{4}{r} T_\theta ^\theta \right]
\label{tena}
\ee
Note that
the assumption that scalar invariants, such as
$T_{MN}T^{MN}$  are finite on the horizon (in order that the latter
is regular),
implies that the boundary terms on the horizon vanish in
(\ref{tena}).
\pr
It is then straightforward to obtain
\be
     (T_r^r)' =\frac{1}{2} \left[ \Gamma' T_t^t
+ \frac{4}{r} T_\theta ^\theta \right]
- \frac{e^{-\frac{\Gamma}{2}}}{r^2} (e^{\frac{\Gamma}{2}} r^2)' T_r^r.
\label{elevena}
\ee
\pr
Ansatz (\ref{twelvea}) for the gauge field yields
\bea
         T_t^t &=& - {\cal E} \nn \\
T_r^r &=& - {\cal E} + {\cal F}  \nn \\
T_\theta ^\theta & = & -{\cal E} + {\cal J}
\label{thirteena}
\eea
with
\bea
{\cal F} &\equiv & \frac{e^{-\Lambda}}{4\pi} \left[ \frac{2\omega '^2}{r^2}
+ \phi '^2 \right]
\nn  \\
{\cal J} &\equiv & \frac{1}{4\pi} \left[ \frac{\omega '}{r^2}e^{-\Lambda}
+ \frac{(1 - \omega ^2)^2}{r^4} + \frac{\phi ^2}{4r^2} (1 + \omega ^2)
\right].
\label{fourteena}
\eea
Substituting (\ref{fourteena}) in (\ref{thirteena}) yields
\be
   T_r^r (r) = \frac{e^{-\frac{\Gamma}{2}}}{r^2}\int _{r_h}^r
\left\{ -
(e^{\frac{\Gamma}{2}}r^2)' {\cal E} + \frac{2}{r} {\cal J}
\right\} dr
\label{fifteena}
\ee
\be
     (T_r^r)' (r) = -\frac{e^{-\frac{\Gamma}{2}}}{r^2}
 (e^{\frac{\Gamma}{2}}r^2)'
{\cal F} + \frac{2}{r}{\cal J}
\label{fifteenb}
\ee
where ${\cal E}$ is expressed as
\be
   {\cal E} =\frac{1}{4\pi} \left[  \frac{(\omega ')^2}{r^2}
e^{-\Lambda } + \frac{(1-\omega ^2)^2}{2r^4} +
\frac{\phi ^2(1 + \omega )^2}{4r^2} + \frac{1}{2} (\phi ')^2e^{-\Lambda }
+ \frac{{\lambda}}{4} (\phi ^2 - { v}^2)^2  \right].
\label{sixteena}
\ee
We now turn to the Einstein equations for the first time, following
the analysis of ref. \cite{bekmod}. Our aim is to examine whether there is a
contradiction with the requirement
of the non-negative energy density. These equations read for our system
\bea
   e^{-\Lambda} (r^{-2} - r^{-1}\Lambda ' ) - r^{-2} &=&
8\pi T_t^t = -8\pi {\cal E} \nn \\
e^{-\Lambda} (r^{-1} \Gamma' + r^{-2} ) - r^{-2} &=& 8\pi T_r^r.
\label{seventeena}
\eea
Integrating out the first of these yields
\be
          e^{-\Lambda }  = 1 -\frac{8\pi}{r} \int _{r_h}^r
{\cal E} r^2 dr - \frac{2 {\cal M_{0}}}{r}
\label{eighteena}
\ee
where ${\cal M_{0}}$ is a constant of integration.
\pr
The requirement for
asymptotic flatness of space-time implies the following
asymptotic behaviour for the energy-density functional ${\cal E} \sim
O(r^{-3}) $ as $r \rightarrow \infty $, so that $\Lambda \sim O(r^{-1}) $.
In order that $e^{\Lambda} \rightarrow \infty $ at the horizon,
$r \rightarrow r_h $, ${\cal M_{0}}$ is fixed by
 \be
               {\cal M_{0}} = \frac{r_h}{2}.
\label{nineteen}
\ee
The second of the equations (\ref{seventeena}) can be rewritten in the form
\be
 e^{-\frac{\Gamma}{2}} r^{-2}
 (r^2 e^{\frac{\Gamma}{2}})' = \left[ 4\pi r T_r^r
+ \frac{1}{2r} \right] e^{\Lambda} + \frac{3}{2r}.
\label{twenty}
\ee
\pr
Consider, first, the behaviour of $T_r^r$ as $ r \rightarrow \infty $.
Asymptotically, $e^{\frac{\Gamma}{2}} \rightarrow 1 $, and so the leading
behaviour of $(T_r^r)'$ is
\be
                   (T_r^r) ' = \frac{2}{r} [{\cal J} - {\cal F} ].
\label{21}
\ee
We, now, note that the fields $\omega $ and $\phi $ have masses
$\frac{{ v}}{2}$ and $\mu  = \sqrt{{ \lambda}}{ v}$
respectively. From the field equations and the requirement of finite energy
density their behaviour at infinity must then be
\bea
      \omega (r) &\sim & -1 + ce^{-\frac{{ v}}{2} r}  \nn \\
      \phi (r)  &\sim & { v} + a e^{-\sqrt{2} \mu r}
\label{22}
\eea
for some constants $c$ and $a$. Hence, the leading asymptotic behaviour
of ${\cal J}$ and ${\cal F}$ is
\bea
  {\cal J} &\sim & \frac{1}{4\pi} \left[ \frac{c^2 { v}^2}{4r^2}e^{-
{ v}r} + \frac{2c^2}{r^4} e^{-{ v}r} +
\frac{{ v}^2 c^2 }{4r^2} e^{-{ v}r} \right] \nn \\
{\cal F} &\sim & \frac{1}{4\pi} \left[ \frac{c^2 { v}^2}
{2r^2} e^{-{ v} r} + 2 a^2 \mu ^2 e^{-\sqrt{2} \mu r} \right]
\label{23}
\eea
since $e^{-\Lambda } \rightarrow 1$ asymptotically.
\pr
The leading behaviour of $(T_r^r)'$, therefore, is
\be
   (T_r^r)' \sim \frac{1}{4\pi} \left[ \frac{2c^2}{r^4} e^{-{ v}r}
- 2a^2 \mu ^2 e^{-2\sqrt{2} \mu r} \right].
\label{24}
\ee
There are two possible cases: (i) $2\sqrt{2}\mu > { v} $ (corresponding
to ${ \lambda } > 1/8$); in this case $(T_r^r)' > 0$
asymptotically, (ii) $2\sqrt{2} \mu \le { v}$ (corresponding to
${ \lambda } \le 1/8$) ; then, $(T_r^r)' < 0$ asymptotically.
\pr
Since ${\cal J}$ vanishes exponentially at infinity, and ${\cal E} \sim
O[r^{-3}]$ as $r \rightarrow \infty$, the integral
defining $T_r^r (r)$ converges as $r \rightarrow \infty $
and $|T_r^r|$ decreases as $r^{-2}$.
\pr
Thus, in case (i) above, $T_r^r$ is negative and increasing as $r \rightarrow
\infty$, and in case (ii) $T_r^r$ is positive and decreasing.
\pr
Now turn to the behaviour of $T_r^r$ at the horizon.
When $r \simeq r_h$ , ${\cal E}$ and ${\cal J}$ are both finite, and
$\Gamma'$ diverges as $1/(r- r_h)$.
Thus the main contribution to $T_r^r$ as $r \simeq r_h$ is
\be
   T_r^r (t) \simeq \frac{e^{-\Gamma/2}}{r^2} \int _{r_h} ^r
(-e^{\Gamma/2}
r^2) \frac{\Gamma'}{2} {\cal E} dr
\label{25}
\ee
which is finite.
\pr
At the horizon, $e^\Gamma = 0$; outside the horizon, $e^\Gamma > 0$ .
Hence $\Gamma' >0$ sufficiently close to the horizon,
 and, since ${\cal E} \ge 0$,
$T_r^r < 0$ for $r$ sufficiently close to the horizon.
\pr
Since ${\cal F} \sim O[r-r_h]$ at $r \simeq r_h$, $(T_r^r)'$ is finite
at the horizon and the leading contribution is
\be
  (T_r^r)' (r_h) \simeq -\frac{\Gamma'}{2} {\cal F} + \frac{2}{r}{\cal J}.
\label{26}
\ee
{}From ref. \cite{greene} we record the relation
\be
re^{-\Lambda} \frac{\Gamma'}{2}   =  e^{-\Lambda}\left[ \omega '^{2} +
\frac{1}{2} r^2 (\phi ')^2 \right] - \frac{1}{2} \frac{(1-\omega ^2)^2}{r^2}
 - \frac{1}{4}\phi ^2 (1 + \omega ^2)^2 + \frac{m}{r} -
\frac{{ \lambda}}{4} (\phi ^2 - { v}^2)^2 r^2
\label{27}
\ee
where $e^{-\Lambda} = 1 - \frac{2m(r)}{r}$. Hence,
\bea
(T_r^r)' &=& -\frac{e^{-\Lambda}}{4\pi r}\left[ \frac{2(\omega ')^2}{r^2}
+ (\phi ')^2 \right] \left\{ (\omega ')^2 + \frac{1}{2} r^2 (\phi ')^2
- \frac{1}{2} e^\Lambda \frac{(1-\omega ^2)}{r^2}  \right.\\
 & & \left.
-\frac{\phi ^2}{4} e^\Lambda (1 + \omega )^2 + \nn
\frac{m}{r}e^\Lambda  \frac{{\lambda}}{4} (\phi ^2 -
{ v}^2 )r^2 e^\Lambda \right\} \\
 & &
+ \frac{1}{2\pi r} \left[ \frac{(\omega ')^2}{r^2} e^\Lambda +
\frac{(1 - \omega ^2)^2}{r^4}  + \frac{\phi ^2}{4r^2}
(1 + \omega)^2 \right].
\label{28}
\eea
For $r \simeq r_h$,
this expression simplifies to
\bea
(T_r^r)'(r_h) & \simeq & {\cal J}(r_h) \left[ \frac{2}{r_h} + \frac{4\pi}{r_h}
{\tilde {\cal F}}(r_h) \right] \nn  \\
 & &  - {\tilde {\cal F}}(r_h) \left[ \frac{1}{2}
+ \frac{1}{2}\frac{(1 - \omega ^2)^2}{r_h^2} -
\frac{{ \lambda}}{4}(\phi ^2 - { v}^2)^2 r_h^2 \right] \nn  \\
 & = &
{\tilde {\cal F}}(r_{h}) \left[
\frac {(1- \omega ^{2})^{2}}{2r_{h}^{3}} +
\frac {\phi ^{2}}{4r_{h}}(1+\omega ^{2})^{2} +
\frac {\lambda }{4}r_{h}(\phi ^{2} -v^{2})^{2} -
\frac {1}{2r_{h}} \right] \nn \\ & &
+ \frac {2}{r_{h}} {\cal J}
\label{29}
\eea
where ${\tilde {\cal F}} = e^\Lambda {\cal F}(r) = \frac{1}{4\pi}
[\frac{2(\omega ')^2}{r^2} + (\phi')^2]$.
\pr
Consider for simplicity the case $r_{h}=1$.
Then, from the field equations \cite{greene}
\bea
\omega _{h}' & = &
\frac {1}{\cal D} \left[
\frac {1}{4} \phi _{h}^{2} (1+\omega _{h}) -
\omega _{h} (1-\omega _{h}^{2}) \right] =
\frac {\cal A}{\cal D} \\
\phi _{h}' & = &
\frac {1}{\cal D} \left[
\frac {1}{2} \phi _{h} (1+\omega _{h})^{2}
+\lambda \phi _{h} (\phi _{h}^{2}- v^{2}) \right] =
\frac {\cal B}{\cal D}
\eea
where
\be
{\cal D}=1- (1-\omega _{h}^{2})^{2} -
\frac {1}{2} \phi _{h}^{2} (1+\omega _{h})^{2}
- \frac {1}{2} \lambda (\phi _{h}^{2} -v^{2})^{2}.
\ee
Then the expression (\ref{29}) becomes
\be
(T^{r}_{r})'(r_{h})=
\frac {1}{4\pi {\cal {D}}} \left[
8\pi {\cal D} {\cal J} - {\cal A}^{2} -\frac {1}{2} {\cal B}^{2}
\right]
=\frac {{\cal {C}}}{ 4  \pi {\cal {D}}}.
\ee
{}From the field equations \cite{greene}
\be
{\cal D}=1-2m_{h}'
\ee
which is always positive because the black holes are
non-extremal~\cite{mw}.
Thus the sign of $(T^{r}_{r})'(r_{h})$ is the same as that of
${\cal C}$.
The EYMH system possesses two branches of solutions~\cite{greene},
labelled by the number of nodes $k$ of the gauge field.
A detailed analysis~\cite{mw} for the case of
black hole solutions possessing at most one node, examined
in refs. \cite{greene,mw},
shows that
for both branches  ($k=1,0$) ${\cal C}$ is {\it  non-negative}.
This implies that
$(T^{r}_{r})'(r_{h})$ is positive for all the black
hole solutions having one node in $\omega $, regardless of the
value of the Higgs mass $v$.
\pr
Let us now check on possible contradictions
with Einstein's equations.
Consider first the case
${ \lambda} > 1/8$. Then, as $r \rightarrow
\infty$, $T_r^r < 0 $ and $(T_r^r)' > 0$.
As $r \rightarrow r_h $, $T_r^r < 0$ and $(T_r^r)' > 0$.
Hence there is no contradiction with
Einstein's equations in this case.
Next,
consider the case $\lambda \le 1/8$.
In this case, as $r \rightarrow \infty$, $T_r^r > 0$ and
$(T_r^r)' < 0$, whilst as $r \rightarrow r_h$, $T_r^r
< 0 $ and $(T_r^r)' > 0$.
Hence, there is an interval $[ r_a, r_b ]$ in which $(T_r^r)' $ is positive
and there exists a `critical' distance $r_c \in (r_a, r_b)$ at which
$T_r^r$ changes sign.
However, unlike the case when the gauge fields are absent~\cite{bekmod},
here
there is {\it no contradiction} with the result following from Einstein
equations, because $(T_r^r)' > 0$ in some open interval close to the
horizon, as we have seen above.
\pr
In conclusion the method of ref. \cite{bekmod} cannot be used to prove
a `no-scalar-hair' theorem for the EYMH system, as expected from the
existence of the
explicit solution of ref. \cite{greene}. The {\it key} difference is
the presence of the positive term $\frac{2}{r}{\cal J}$
in the expression (\ref{fifteenb}) for $(T_r^r)'$. This term is
dependent on the Yang-Mills field and {\it vanishes} if this field
is {\it absent}, or if the field is {\it Abelian}.
Thus, there is a sort of `balancing' between the gravitational
attraction and the {\it non-Abelian} gauge field repulsion, which
is responsible for the existence of the classical non-trivial
black-hole solution of ref. \cite{greene}.
However, as shown in ref. \cite{mw}, this solution is not stable against
(linear) perturbations of the various field configurations.
Thus, although the `letter' of the `no-scalar-hair' theorem of ref.
\cite{bekmod}, based on non-negative scalar-field-energy density,
is violated, its `spirit' is maintained in the sense that there exist
instabilities which imply that the solution cannot be formed
as a result of collapse of stable matter.
However, stability is a new dimension in the
no-hair conjecture, not included in the original
formulation. Therefore,  it seems fair to
say that the above analysis constitutes an analytic proof
of  bypassing or, better,  `eluding' the no-hair conjecture.
\pr
\section{Dilatonic Hair in Higher-Curvature Gravity}
\pr
As a second example of a physical theory of black holes
not covered by the no-hair theorem of ref. \cite{bekmod}
I shall describe
black hole
solutions~\cite{rizos}
of the Einstein-dilaton system in the presence of the
(higher-derivative) curvature-squared terms
of Gauss-Bonnet (GB) type~\cite{rizos}.
These solutions were found numerically by P. Kanti, J. Rizos and
K. Tamvakis, and
are
endowed with a non-trivial dilaton field outside the horizon, thus
possessing dilaton hair. The treatment of the curvature-squared
terms in ref. \cite{rizos} is
non-perturbative and the solutions are present for any value of
$\alpha'/g^2$, where $\alpha '$ is the string Regge slope,
and $g$ is the gauge coupling constant of the low-energy theory.
What I shall argue in this section, in connection with a
bypassing of the no-hair theorem of ref. \cite{bekmod},
is that the presence of the higher-derivative GB terms
provides the necessary `repulsion' in the effective
theory that balances the gravitational
attraction, thereby leading to
black holes dressed with non-trivial classical
dilaton hair. This is
an analogous phenomenon to that occuring
in the case
of Einstein-Yang-Mills systems discussed in the previous section.
\paragraph{}
The action I shall use will be the effective low-energy
action obtained from (super)strings.
I shall concentrate on the bosonic part of the
gravitational multiplet
which consists of the dilaton, graviton,
and antisymmetric tensor fields. I shall ignore
the antisymmetric tensor for simplicity\footnote{In four dimensions,
the antisymmetric tensor field leads to the axion hair, already
discussed in ref.\cite{kanti}. Modulo unexpected surprises,
we do not envisage problems
associated with its presence as regards
the results discussed in this work, and, hence,
we ignore it for simplicity.}.
 As is well known in low-energy effective field theory,
there are ambiguities in the coefficients of such terms,
due to the possibility of {\it local} field redefinitions which
leave the $S$-matrix amplitudes of the effective
field theory invariant,
according to the {\it equivalence} theorem.
To ${\cal O}(\alpha ')$ the freedom of such redefinitions
is restricted to two generic structures, which cannot be removed by
further redefinitions~\cite{metsaev}. One is a
 curvature-squared combination,
and the other is a four-derivative dilaton term.
Thus, a generic form of the string-inspired  ${\cal O}(\alpha ')$
corrections to Einstein's gravitation have the form
\begin{equation}
{\cal L}=-\frac{1}{2}R - \frac{1}{4} (\partial _\mu \phi )^2
+ \frac{\alpha '}{8g^2} e^{\phi } (c_1{\cal R}^2  + c_2
(\partial _\rho \phi )^4 )
\label{one}
\end{equation}
where $\alpha '$ is the Regge slope, $g^2$
is some gauge coupling constant (in the case
of the heterotic string that we concentrate
for physical reasons), and
${\cal R}^2$ is a generic curvature-dependent
quadratic  structure, which can always be fixed
to correspond to the Gauss-Bonnet (GB) invariant
\begin{equation}
R^2_{GB} =
R_{\mu\nu\rho\sigma}R^{\mu\nu\rho\sigma} -
4 R_{\mu\nu}R^{\mu\nu} + R^2
\label{two}
\end{equation}
The coefficients $c_1$, $c_2$ are fixed by comparison
with string scattering amplitude computations,
or $\sigma$-model $\beta$-function analysis. It is known that
in the three types of string theories, Bosonic, Closed-Type II Superstring,
and Heterotic Strings, the ratio of the $c_1$ coefficients
is 2:0:1 respectively~\cite{metsaev}.
The case of superstring II effective theory, then,
is characterized by the absence of curvature-squared terms.
In such theories the fourth-order dilaton terms can still be, and in
fact they are, present. In such a case, it is straightforward to
see from the modern proof of the no-scalar hair theorem
of ref. \cite{bekmod} that such theories, cannot sustain to
order ${\cal O}(\alpha ')$, any non-trivial dilaton hair.
On the other hand, the presence of curvature-squared terms
can drastically change the situation~\cite{rizos}, as I
will now describe.
\paragraph{}
Following the above discussion we shall ignore, for simplicity,
the fourth-derivative dilaton terms
in (\ref{one}), setting from now on $c_2=0$.
However, we must always bear in mind that such terms are non-zero
in realistic effective string cases, once the GB combination
is fixed for the gravitational ${\cal O}(\alpha ')$ parts.
Then, the lagrangian for dilaton gravity with a
Gauss Bonnet term reads
\begin{equation}
{\cal L}=-\frac{1}{2}R - \frac{1}{4} (\partial _\mu \phi )^2
+ \frac{\alpha '}{8g^2} e^{\phi } R^2 _{GB}
\label{three}
\end{equation}
where  $R^2_{GB}$ is the Gauss Bonnet (GB) term (\ref{two}).
\paragraph{}
As I  mentioned earlier, although we view (\ref{three})
as a heterotic-string effective action, for simplicity, in this paper
we shall ignore the modulus and axion
fields, assuming reality of the dilaton ($S=e^{\phi}$ in
the notation of ref. \cite{kanti}).
We commence our analysis by noting that
the dilaton field and Einstein's equations derived from (\ref{three}),
are
\begin{eqnarray}
&~&\frac{1}{\sqrt{-g}} \partial _\mu [\sqrt{-g} \partial ^\mu \phi ]
=-\frac{\alpha '}{4g^2} e^\phi R^2_{GB}
\label{fourab} \\[3mm]
&~&R_{\mu\nu} - \frac{1}{2} g_{\mu\nu} R  =
- \frac{1}{2} \partial _\mu \phi
\partial _\nu \phi + \frac{1}{4} g_{\mu\nu} (\partial _\rho \phi )^2  -
\alpha ' {\cal K}_{\mu\nu}
\label{fourb}
\end{eqnarray}
where
\begin{equation}
{\cal K}_{\mu\nu}=(g_{\mu\rho}g_{\nu\lambda}+g_{\mu\lambda}g_{\nu\rho})
\eta^{\kappa\lambda\alpha\beta} D _\gamma
[{\tilde R}^{\rho\gamma}_{\,\,\,\,\,\alpha\beta} \partial _\kappa f]
\end{equation}
and
\begin{eqnarray}
\eta ^{\mu\nu\rho\sigma} &=& \epsilon ^{\mu\nu\rho\sigma}
(-g)^{-\frac{1}{2}}\nonumber \\[2mm]
\epsilon ^{0ijk} &=& -\epsilon_{ijk} \nonumber \\[2mm]
{\tilde R}^{\mu\nu}_{\,\,\,\,\,\kappa\lambda} &=& \eta^{\mu\nu\rho\sigma}
R_{\rho\sigma\kappa\lambda}  \\[2mm]
f & = & \frac {e^\phi} {8 g^2}\nonumber
\end{eqnarray}
{}From the right-hand-side
of the modified Einstein's equation (\ref{fourb}),
one can construct a conserved
``energy momentum tensor'', $D _\mu T^{\mu\nu} = 0$,
\begin{eqnarray}
T_{\mu\nu} =  \frac{1}{2} \partial_\mu\phi \partial_\nu \phi - \frac{1}{4}
g_{\mu\nu} (\partial _\rho \phi )^2  + \alpha ' {\cal K}_{\mu\nu}
\label{five}
\end{eqnarray}
It should be stressed that the time component of $-T_{\mu\nu}$,
which in Einstein's gravity would correspond to the
local energy
density ${\cal E}$, may {\it not be positive }.
Indeed, as we shall see later on,
for spherically-symmetric space times, there are regions
where this quantity is negative.
The reason is that, as a result of the higher derivative GB terms,
there are contributions of the gravitational field itself
to $T_{\mu\nu}$. From a string theory point of view, this is
reflected in the fact that the dilaton is part of the string
gravitational multiplet.
Thus, this is the {\it first} important indication
on the possibility of evading the no-scalar-hair theorem
of ref. \cite{bekmod} in this case.
However, this by itself is not
sufficient for a rigorous proof of an
evasion of the no-hair conjecture.
I  shall come to this point later on.
\paragraph{}
At the moment, let me
consider a spherically symmetric space-time having the metric
\begin{equation}
ds^2 = -e^\Gamma dt^2 + e^\Lambda dr^2 + r^2 (d\theta ^2 + sin^2 \theta
d\varphi^2)
\label{six}
\end{equation}
where $\Gamma$, $\Lambda$ depend on $r$ solely. Using the above ansatz,
the dilaton equation as well as the $(tt)$, $(rr)$ and $(\theta\theta)$
component of the Einstein's equations take the form
\begin{eqnarray}
&~&\hspace{-0.7cm} \phi''+\phi'(\frac{\Gamma'-\Lambda'}{2}+\frac{2}{r})=
\frac{\alpha'e^\phi}{g^2 r^2}\left(\Gamma'\Lambda'e^{-\Lambda}+
(1-e^{-\Lambda})[\Gamma''+
\frac{\Gamma'}{2}(\Gamma'-\Lambda')]\right)
\label{sixteen} \\[3mm]
&~&\hspace{-0.7cm} \Lambda'\left(1+\frac {\alpha' e^\phi} {2 g^2 r} \phi'
(1-3 e^{-\Lambda})\right)
=\frac{r\phi'^2}{4}+\frac {1-e^\Lambda}{r}
+\frac{\alpha' e^\phi}{g^2 r}(\phi''+\phi'^2)(1-e^{-\Lambda})
\label{seventeen} \\ [3mm]
&~&\hspace{-0.7cm} \Gamma'\left(1+\frac {\alpha' e^\phi}{2 g^2 r}\phi'
(1-3 e^{-\Lambda})\right)
=\frac{r \phi'^2}{4}+\frac{e^\Lambda-1}{r}
\label{eighteen} \\[3mm]
&~&\hspace{-0.7cm} \Gamma''+\frac{\Gamma'}{2}(\Gamma'-\Lambda')+
\frac{\Gamma'-\Lambda'}{r}
=-\frac{{\phi'}^2}{2}+\frac{\alpha' e^{\phi-\Lambda}}{g^2 r}
\left(\phi'\Gamma''+(\phi''+{\phi'}^2)\Gamma'\right.\nonumber \\[3mm]
&~&\hspace{5.4cm}\left. +\frac{\phi'\Gamma'}{2}(\Gamma'-3\Lambda')\right)
\label{nineteenb}
\end{eqnarray}
Before I proceed to study the above system it is useful to note that
if I turn off the Gauss-Bonnet term, equation (10) can be
integrated to give $\phi'\sim \frac{1}{r^2} e^{(\Lambda-\Gamma)/2}$.
A black hole solution should have at the horizon $r_h$ the behaviour
$e^{-\Gamma}$, $e^\Lambda \rightarrow \infty$. Therefore the radial
derivative of the dilaton would diverge on the horizon resulting into
a divergent energy-momentum tensor
\begin{equation}
T^t_t=-T^r_r= T^\theta _\theta=-\frac{e^{-\Lambda}}{4} \phi'^2
\rightarrow \infty
\end{equation}
Rejecting this solution we are left with the standard Schwarzschild
solution and a trivial $(\phi=constant)$ dilaton, in agreement with
the no-hair theorem. This behaviour will be drastically modified
by the Gauss-Bonnet term.
\paragraph{}
The $r$ component of the energy-momentum
conservation equations reads:
\begin{equation}
(e^{\Gamma /2}r^2 T_r^r )' =\frac{1}{2} e^{\Gamma /2} r^2
[\Gamma ' T_t^t + \frac{4}{r}T_\theta^\theta]
\end{equation}
where the prime denotes differentiation with respect to $r$. The
spherical symmetry of the space-time implies $T_\theta^\theta =
T_\varphi^\varphi $. Integrating over the radial coordinate $r$
from the horizon $r_h$ to generic $r$ yields
\begin{equation}
T_r^r (r)  =  \frac{e^{-\Gamma /2}}{2r^2} \int _{r_h}^r
e^{\Gamma /2} r^2 [\Gamma ' T_t^t + \frac{4}{r}
T_\theta ^\theta ] dr
\label{eight}
\end{equation}
The boundary terms on the horizon vanish, since scalar invariants
such as $T_{\alpha\beta}T^{\alpha\beta}$ are finite there.
For the first derivative of $T_r^r$ we have
\begin{equation}
(T_r^r)'(r) = \frac{e^{-\Gamma /2}}{r^2} (e^{\Gamma /2} r^2)'
\,(T^t_t - T^r_r) + \frac{2}{r}\,(T^{\theta}_{\theta} - T^r_r)
\label{nine}
\end{equation}
Taking into account (\ref{five}) and (\ref{six}), one easily obtains
\begin{eqnarray}
T^t_t&=&- e^{-\Lambda} \frac {\phi'^2}{4}-
\frac {\alpha'}{g^2 r^2}e^{\phi-\Lambda} (\phi''+\phi'^2)
 (1-e^{-\Lambda}) + \frac {\alpha'} {2 g^2 r^2} e^{\phi-\Lambda}
 \phi' \Lambda' (1-3 e^{-\Lambda}) \nonumber \\ [3mm]
T_r^r&=& e^{-\Lambda} \frac {\phi'^2} {4}-
\frac {\alpha'} {2 g^2 r^2} e^{\phi-\Lambda} \phi'
 \Gamma' (1-3 e^{-\Lambda})  \label{comp} \\ [3mm]
T_\theta^\theta&=&- e^{-\Lambda} \frac {\phi'^2}{4}
 +\frac {\alpha '}{2 g^2 r} e^{\phi-2 \Lambda}
 [\Gamma'' \phi' + \Gamma' (\phi'' + \phi'^2) + \frac {\Gamma'
 \phi'} {2} (\Gamma'-3 \Lambda')] \nonumber
\end{eqnarray}
In the relations (\ref{comp}) there lies the {\it second} reason
for a possibility of an evasion of the no-hair conjecture.
Due to the presence
of the higher curvature contributions, the
relation $T_t^t = T_\theta^\theta$
assumed in ref. \cite{bekmod}, is no longer valid.
The alert reader must have noticed, then,
the similarity of the r\^ole played by the Gauss-Bonnet
${\cal O}(\alpha ')$ terms in the lagrangian (\ref{three})
with the case of the non-Abelian gauge black holes
studied in ref. \cite{mw}, and described in the previous
section.
We stress once, again, however, that in the GB case
{\it both} the non-positivity of the ``energy-density'' $T_t^t$,
and the modified relation $T_t^t \ne T_\theta^\theta$,
play equally important r\^oles in leading to a possibility
of having non-trivial classical scalar (dilaton)
hair in GB black holes systems.
Below I shall demonstrate rigorously this, by showing that
there is {\it no contradiction} between the
results following from the conservation
equation of the ``energy-momentum tensor'' $T_{\mu\nu}$ and
the field equations, in the presence of non trivial dilaton hair.
\paragraph{}
First, let me define what one means by `dilaton hair'.
Far away from the origin the unknown functions $\phi(r)$, $e^{\Lambda(r)}$,
and $e^{\Gamma(r)}$ can be expanded in a power series in $1/r$. These
expansions, substituted back into the equations, are finally expressed
in terms of three parameters only, chosen to be $\phi_{\infty}$,
the asymptotic value of the dilaton, the ADM mass $M$, and the
dilaton charge
$D$ defined as~\cite{mitra}
\begin{equation}
D=-\frac{1}{4\pi}\int d^2\Sigma^\mu D_\mu \phi
\end{equation}
where the integral is over a two-sphere at spatial infinity.
The asymptotic solutions are
\begin{eqnarray}
e^{\Lambda(r)}&=&1+\frac{2 M}{r}+ \frac{16 M^2-D^2}{4 r^2}
+{\cal O}(1/r^3)
\label{thirteenab} \\ [3mm]
e^{\Gamma(r)}&=&1-\frac{2 M}{r} + {\cal O}(1/r^3)
\label{thirteenb} \\ [3mm]
\phi(r)&=& \phi_{\infty}+\frac{D}{r}+\frac{M D}{r^2}
+ {\cal O}(1/r^3)
\label{thirteenc}
\end{eqnarray}
\pr
Note, now, that
the  solution near the horizon is
characterized by the parameter $\phi_h$ .
However, the  parameters that characterize
the solution near infinity (\ref{thirteenab})-(\ref{thirteenc})
are  $M$ and $D$ . From this, we can infer that a relation
must hold between the
above parameters in order to be able to classify the solution
as a
one parameter family of black hole solutions. After some
manipulation,
the set of equations (\ref{sixteen})-(\ref{nineteenb}) can
be rearranged to yield the identity
\begin{equation}
\frac{d}{dr}\left(r^2 e^{(\Gamma-\Lambda)/2}(\Gamma'-\phi')-
\frac{\alpha' e^\phi}{g^2}
e^{(\Gamma-\Lambda)/2} [(1-e^{-\Lambda}) (\phi'-\Gamma')+
e^{-\Lambda} r \phi' \Gamma']\right)=0
\end{equation}
Integrating this relation over the interval $(r_h,r)$ we obtain
the expression
\begin{equation}
 2 M -D=\sqrt{\gamma_1 \lambda_1} (r_h^2 +\frac{\alpha' e^{\phi_h}}
 {g^2})
\end{equation}
This equation is simply a connection between the set of parameters
describing the solution near the horizon and the set $M$ and $D$.
The rhs of this relation clearly indicates that the existing dependence
of the dilaton charge on the mass does not take the simple form of an
equality encountered in EYMD regular solutions of ref.\cite{donets}.
In order to find the relation between $M$ and $D$ we follow
refs.\cite{kanti}
and take into account the ${\cal O}(\alpha^{'2})$
expression of the dilaton charge in the limit $r \rightarrow \infty$
\begin{eqnarray}
\phi(r)&=&\phi_\infty + \frac{D}{r} +... \nonumber \\ [3mm]
&=&\phi_{\infty} + \left(\frac{e^{\phi_\infty}}{2 M}
 \frac{\alpha'}{g^2} + \frac{73 e^{2 \phi_\infty}}{60 (2 M)^3}
 \frac {\alpha^{'2}}{g^4}\right) \frac{1}{r} +...
\label{dm}
\end{eqnarray}
This relation can be checked numerically for the
black hole solution of ref.~\cite{rizos}.
Any deviations from this relation
are due to higher order terms which turn out to be small.
The above relation (\ref{dm}) implies that the dilaton hair
of the black hole solution, if exists, is a kind
of `secondary hair', in the terminology of ref. \cite{coleman}.
This hair is generated because the basic fields  (gravitons) of
the theory associated
with the primary hair (mass) act as sources for the non-trivial
dilaton
configurations outside the horizon of the black hole.
\pr
To check the possibility of the evasion of the no-hair conjecture
we first consider the asymptotic behaviour of $T_r^r$ as
$r \rightarrow \infty$. Since $\Gamma '$ and $\Lambda '$
$\sim {\cal O}(\frac{1}{r^2}) $ as $r \rightarrow \infty$,
we have the following asymptotic behaviour
\begin{eqnarray}
T^r_r & \sim & \frac{1}{4} (\phi ')^2 + {\cal O}(\frac{1}{r^6})
\nonumber \\ [3mm]
T^\theta_\theta &\sim &-\frac{1}{4} (\phi ')^2 + {\cal O}(\frac{1}{r^6})
\end{eqnarray}
In this limit, $e^{\Gamma/2} \rightarrow 1$, and so the leading behaviour
of $(T_r^r)'$ is
\begin{equation}
(T_r^r)' \sim \frac{2}{r} (T^\theta_\theta - T^r_r) \sim -\frac{1}{r}
(\phi ')^2 < 0 \qquad {\rm as~r~\rightarrow \infty}
\end{equation}
Thus, $T_r^r$ is positive and decreasing as $r \rightarrow \infty$.
\paragraph{}
We now turn to the behaviour of the unknown functions at the event horizon.
When $r \sim r_h$, we make the ansatz
\begin{eqnarray}
e^{-\Lambda(r)}&=&\lambda_1 (r-r_h) + \lambda_2 (r-r_h)^2 +...
\nonumber \\[3mm]
e^{\Gamma(r)}&=&\gamma_1 (r-r_h) + \gamma_2 (r-r_h)^2 +...
\label{fourteen} \\[3mm]
\phi(r)&=& \phi_h + \phi'_h (r-r_h) + \phi''_h(r-r_h)^2 +...
\nonumber
\end{eqnarray}
with the subscript $h$ denoting the value of the respective
quantities at the horizon.
It can be shown~\cite{rizos} that this is the most general
asymptotic solution $\Gamma'\rightarrow\infty ,\phi,\phi'$ finite.
As we can see, $\phi(r_h)\sim constant$ while $\Gamma '$ and $\Lambda '$
diverge as $(r-r_h)^{-1}$ and $-(r-r_h)^{-1}$ respectively. Then, the
behaviour of the components of
the energy-momentum tensor near the horizon is
\begin{eqnarray}
T_r^r & = & -\frac {\alpha'} {2 g^2 r^2} e^{\phi-\Lambda} \phi'
\Gamma' + {\cal O}(r-r_h) \nonumber \\ [3mm]
T_t^t & = & \frac {\alpha'} {2 g^2 r^2} e^{\phi-\Lambda} \phi'
\Lambda' + {\cal O}(r-r_h)  \\ [3mm]
T_\theta^\theta &=& \frac {\alpha'} {2 g^2 r} e^{\phi-2 \Lambda}
[\Gamma'' \phi'+ \frac {\Gamma' \phi'} {2} (\Gamma'-3 \Lambda')] +
{\cal O}(r-r_h) \nonumber
\label{fifteen}
\end{eqnarray}
Taking into account the above expressions the leading behaviour
of $T_r^r$ near the horizon is
\begin{equation}
T_r^r (r) \sim -\frac{e^{-\Gamma /2}}{r^2} \int _{r_h}^r
\frac{\alpha '}{4g^2}e^{\Gamma/2} (\Gamma ')^2  e^{-\Lambda} e^\phi
\phi ' dr + {\cal O}(r-r_h)
\end{equation}
Therefore one observes that
for $r$ sufficiently close to the event horizon, $T_r^r$
has {\it opposite} sign to $\phi '$.
\paragraph{}
For $(T_r^r)'$ near the horizon,
we have
\begin{eqnarray}
(T_r^r)'(r)&=& \frac{\alpha '}{2g^2} \frac{e^\phi}{r^2} e^{-\Lambda}
\{ -\Gamma ' (\phi '' + \phi'^2) +
\phi ' [ \frac{\Gamma '}{2} (\Gamma ' + \Lambda ') +
2 e^{-\Lambda} \Gamma '' - \frac{2}{r} \Lambda ' ] \}\nonumber \\[3mm]
&~&-\frac{1}{4} \Gamma ' e^{-\Lambda} \phi'^2+ {\cal O}(r-r_h)
\label{fifteenab}
\end{eqnarray}
where $\Gamma ' + \Lambda ' \sim {\cal O}(1)$ for $r \sim r_h$.
Adding the $(tt)$ and $(rr)$ components of the Einstein's equations
we obtain at the event horizon
\begin{equation}
\Gamma ' + \Lambda ' = \frac{1}{{\cal F}} [ \frac{1}{2}
r_h \phi_h'^2 + \frac{\alpha '}{g^2} \frac{e^{\phi _h}}{r_h}
(\phi_h'' + \phi_h'^2 ) ] + {\cal O}(r-r_h)
\end{equation}
where
\begin{equation}
{\cal F} = 1 + \frac{\alpha '}{2g^2} \frac{e^{\phi _h}}{r_h} \phi'_h
\label{thirtyb}
\end{equation}
{}From the $(\theta\theta)$ component we obtain
\begin{eqnarray}
e^{-2\Lambda} \Gamma '' &=& -\frac{1}{2} e^{-2\Lambda}
(\Gamma ')^2 + \frac{1}{2} e^{-2\Lambda} \Gamma ' \Lambda '
+ {\cal O}(r-r_h)  \nonumber \\[2mm]
&=&-\frac{1}{r_h^2 {\cal F}^2 } + {\cal O}(r-r_h)
\label{twentyb}
\end{eqnarray}
Substituting all the above formulae into (\ref{fifteenab})
yields, near $r_h$
\begin{equation}
(T_r^r)' (r) \sim -\frac{1}{4}\frac{\phi_h'^2}{r_h^2 {\cal F}}
- \frac{\alpha '}{2g^2} \frac{e^{\phi _h}}{r_h^3 {\cal F}^2 }
(\phi_h'' + \phi_h'^2) - \frac{\alpha '}{4g^4}
\frac{e^{2\phi _h}}{r^5_h {\cal F}^2} \phi_h'^2 + {\cal O}(r-r_h)
\label{twentyc}
\end{equation}
\paragraph{}
Next, we turn to the dilaton equation (\ref{sixteen}). At $r \sim r_h$,
it takes the form
\begin{equation}
\frac{\phi _h'}{r_h {\cal F}} = - \frac{3}{{\cal F}^2}
\frac{\alpha '}{g^2} \frac{e^{\phi _h}}{r_h^4} + {\cal O}(r-r_h)
\end{equation}
Substituting for ${\cal F}$ (\ref{thirtyb}), the following equation
for $\phi_h '$ is derived
\begin{equation}
\frac{\alpha '}{2g^2} \frac{e^{\phi _h}}{r_h}
\phi_h'^2+ \phi'_h+ \frac{3}{r_h^3} \frac{\alpha '}{g^2}
e^{\phi _h} = 0
\end{equation}
which has as solutions
\begin{equation}
\phi _h' = \frac{g^2}{\alpha '}r_h e^{-\phi _h}\left(
-1 \pm \sqrt{1 - \frac{6(\alpha ')^2  }{g^4}
\frac{e^{2\phi _h}}{r_h^4}}\right)
\label{twentyfive}
\end{equation}
One can show~\cite{rizos} that
the relation (\ref{twentyfive}) guarantees
the {\it finiteness} of $\phi _h ''$, and hence of
the ``local density'' $T_t^t$ (\ref{comp}).
Both these solutions for $\phi_h'$ are negative, and hence, since
$T_r^r(r_h)$  has the opposite sign to $\phi _h'$, $T_r^r$
will be {\it positive} sufficiently close to the horizon.
Since $T_r^r \ge 0$ also at infinity, we observe that there is
{\it no contradiction} with Einstein's equations, thereby
allowing for the existence of black holes with scalar hair.
We observe that near the horizon the quantity
${\cal E}$ ($-T_t^t$), which in Einstein's
gravitation would be the local energy density
of the field $\phi$, is
{\it negative}. As we mentioned earlier, this
constitutes one of the reasons one should expect
an evasion of the no-scalar-hair conjecture
in this black hole space time.
Crucial also for this result was the presence of
additional terms in (\ref{comp}),
leading to $T_t^t \ne T_\theta^\theta$.  Both of these
features, whose absence in
the case of Einstein-scalar gravity was
crucial for the modern proof of the no-hair theorem,
owe their existence in the presence
of the higher-order ${\cal O}(\alpha ')$ corrections
in (\ref{three}).
\paragraph{}
The physical importance of the restriction (\ref{twentyfive})
lies on the fact that according to this relation,
black hole solutions of a given horizon radius
can {\it only exist} if the coupling constant of the
Gauss-Bonnet term in (\ref{three}) is smaller than
a critical value, set by the magnitude of
the horizon scale. In fact from (\ref{twentyfive}), reality
of $\phi _h'$ is guaranteed if and only if
\begin{equation}
    e^{\phi _h} < \frac{g^2}{\sqrt{6}\alpha '} r_h^2
\label{tsix}
\end{equation}
In this picture, $\beta \equiv \frac{1}{4}e^{\phi _h}$ can
then be viewed as
the (appropriately normalized with respect to the Einstein term)
coupling constant of the GB term in the effective lagrangian
(\ref{three}). For a black hole of unit horizon radius $r_h =1$,
the critical value of $\beta$, above which black hole solutions
cannot exist, is then $\beta _c = g^2/4\sqrt{6}\alpha'$.
One is tempted to compare the situation with the case of $SU(2)$
sphaleron solutions in the presence of Gauss Bonnet
terms \cite{donets}.
Numerical analysis of sphaleron solutions in such systems reveals
the existence of a critical value for the GB coefficient
above which solutions do not exist. In the sphaleron case
this number depends on the number of nodes of the
Yang-Mills gauge field. In our case, if
one fixes the position
of the horizon, then it seems that in order to
construct black hole solutions with this horizon size
the GB coefficient has to satisfy (\ref{tsix}).
Thus, a way of interpreting (\ref{twentyfive}) is to
view it as
providing a necessary condition for the {\it absence of naked
singularities} in space-time~\cite{rizos}.
To understand better this latter point we should consider the
scalar curvature in the limit $r \rightarrow r_h$. It is
given by
\begin{equation}
R=\frac{2}{r_h^2}\,\left(\frac{1 \mp \sqrt{1 - \frac{6(\alpha ')^2  }{g^4}
\frac{e^{2\phi_h}}{r_h^4}}}{1 \pm \sqrt{1 - \frac{6(\alpha ')^2  }{g^4}
\frac{e^{2\phi_h}}{r_h^4}}}\right)^2
\end{equation}
This expression shows that the curvature is singular at
$r_h \rightarrow 0$, i.e. when the horizon shrinks. The point
$r_h=0$ can be reached only when $\phi_h=-\infty$. Thus
the inequality (36), in a sense, forbids $r_h$ to become zero
and reveal the singularity.
\paragraph{}
Above, we have argued on the possibility of
having black holes in the system (\ref{three}) that admit
non-trivial dilaton hair outside their horizon.
The key is the bypassing of the no-hair theorem~\cite{bekmod}, as a result of
the curvature-squared terms. However, the hair
appears to be of secondary type, not yielding new quantum
numbers for the black hole, but expressed in terms of its
$ADM$
mass.
In ref. \cite{rizos} Kanti et al.
found
explicit black-hole solutions
of the equations of motion originating from
(\ref{three}) and provided evidence for the existence
of black hole solutions to all orders in $\alpha '$.
Unfortunately a complete analytic treatment
of these equations is not feasible, and one has to
use numerical methods. This complicates certain things,
in particular it does not allow for a clear view of
what happens inside the horizon, thereby not giving
good information on the curvature singularity structure.
Nevertheless, the existence of an horizon in those
solutions is demonstrated~\cite{rizos}, and, thus, the
GB-dilaton system may be considered as
constituting a second example
of the `elusion' of the no-hair conjecture.

\section{Conclusions}
\paragraph{}
In this paper
I have discussed
two quite generic examples
of bypassing the no-scalar-hair conjecture
for black holes: the Einstein-Yang-Mills-Higgs $SU(2)$
system~\cite{greene,mw},
possessing black holes with non-trivial Higgs hair
outside the horizon, and the Gauss-Bonnet (GB)
higher-curvature graviton-dilaton
system, possessing black holes
with dilaton hair~\cite{rizos}.
\pr
I have presented a method
of proving analytically the existence of scalar hair for such
systems~\cite{mw}. I believe
this is of value, not only because
the black hole solutions in both systems are known only
numerically at present, but also because
the above examples
constitute two rather generic categories
of black hole systems that may evade the no-hair
conjecture.
The physical origin behind the existence
of the hairy black holes in both systems
can be traced back to
the existence of non-Abelian field repulsion:
the repulsion due to the Yang-Mills gauge field in the EYMH case,
or the repulsion due to the presence of the higher-curvature
(gravitational) terms in the GB system. Such repulsion
balances the gravitational attraction from the Einstein terms,
leading to non-trivial black-hole space-time structure.
\pr
Although the `letter' of the no-scalar-hair theorem
is violated in both cases, however its `spirit'
remains valid since the black hole solutions
in the EYMH system are unstable~\cite{mw,winst},
and the dilaton hair in the GB
is of `secondary' type~\footnote{The
stability of the GB-dilaton system has not been studied yet.},
in the
sense that it
is not accompanied by the
presence of any new quantity that characterizes the black
hole
given that
the dilaton charge
can be
expressed in terms of the $ADM$ mass of the black hole.
It should be
stressed, however,
that irrespectively of the precise type of hair
the set of solutions examined in this talk
bypasses the conditions of the no-scalar-hair theorem~\cite{bekmod}.
Thus, our work~\cite{mw,rizos} may be viewed as
demonstrating that there is plenty
of room in the gravitational
structure of Non-Abelian Gauge and/or
Superstring Theory to allow for
physically sensible situations that are not covered by the
no-hair theorem as stated.
\pr
The relevance of the above
evasion of the no-hair conjecture to the information-loss
and quantum decoherence
issues of a black-hole space time
remains to be seen. Although the analytic proof allows the existence
of hair, it does not provide any information on the amount
of hair carried by the black holes, or on whether
such hair is capable of storing enough information
so as to maintain unitarity of the `black-hole plus matter'
system during Hawking evaporation. 
It seems more likely that quantum 
hair, if exists,will be more 
relevant for this purpose. Such questions,
especially from the point of view of string-inspired
theories~\cite{emn},
are left
for future investigations.

\section*{Acknowledgments}
It is a pleasure to thank the organizers of the 5th Hellenic
School and Workshops on HEP at Corfu, for providing
a very stimulating environment for research
and discussions
during the meeting.
The above talk was based on
work
done
in collaboration with P. Kanti, J. Rizos,
K. Tamvakis and E. Winstanley.
 

%
\end{document}